\documentstyle[twocolumn,psfig,prl,aps]{revtex}
\begin{document} 
\twocolumn[\hsize\textwidth\columnwidth\hsize\csname @twocolumnfalse\endcsname

\title{Tunneling and orthogonality catastrophe in the topological
mechanism of superconductivity}

\author{A.G. Abanov, P.B. Wiegmann}
\address{James Franck Institute of the University of Chicago,
5640 S.Ellis Avenue, Chicago, IL 60637, USA}
\maketitle

\begin{abstract}
We compute the angular dependence of the order parameter and tunneling
amplitude in a model exhibiting topological superconductivity and sketch its
derivation as a model of a doped Mott insulator. We show that ground states
differing by an odd number of particles are orthogonal and the order parameter
is in the d-representation, although the gap in the electronic spectrum has no nodes.
We also develop an operator algebra, that allows one to compute off-diagonal
correlation functions. \\
\end{abstract}]

1. In the BCS theory of superconductivity
numerous physical quantities like London's penetration depth, the gap in the
electronic spectrum, the tunneling amplitude, etc. are expressed through a single
object - an off-diagonal two-particle matrix element between ground states with $N$
and $N+2$ particles
\begin{equation}
\label{dts}
\Delta(r-r')=\varepsilon^{\sigma\sigma'}\langle N|c_{\sigma}(r)c_{\sigma'}(r')|N+2\rangle 
\end{equation}
This fact is not special to superconductivity, but rather a
manifestation of the mean-field character of BCS. In electronic liquids
where the interaction is strong in the high temperature regime above  $T_c$, one
expects also to see a difference between these different implementations of 
superconductivity.

Below we consider the extreme case of a strongly interacting system where a ground state
(and the entire spectrum) depends crucially on the number of electrons in the system. In
recent years it has been argued that under rather general assumptions, an electronic
system where a topological soliton is adiabatically attached to a particle develops
superconductivity.  Realizations of this phenomenon  in low dimensions are Fr\"ohlich
ideal conductivity in 1D \cite{Frohlich} and  anyon superconductivity in 2D
\cite{anyonSup}. We refer to this phenomenon as  topological superconductivity
\cite{WTS1,WTS2}. The crucial features of the mechanism are (i) an electron acquires a
geometrical phase  in  field of a soliton, and a related
phenomenon: (ii) orthogonality catastrophe. Both of them make the physics of the
superconducting state drastically different from BCS physics and in particular give
rise to an angle dependence of the matrix element (\ref{dts}) and the tunneling amplitude.

In this Letter we consider an ideal 2D model of a doped Mott insulator on a
square lattice at doping close to the half filling. In this model the
Fermi surface consists of four pockets around $k_f=(\pm\pi/2,\pm\pi/2)$. We
calculate the Josephson tunneling amplitude \cite{RL} and show that the phase
difference between the tunneling amplitude on the different faces (1,0) and (0,1)
of the crystal is $\pi$. At the same time the gap has no nodes. This result is
depicted in Fig.1 and is expressed by eq. (\ref{kl}).  Although it is
in agreement with the corner-SQUID-junction experiment \cite{Jos} the order parameter 
is different from the conventional d-wave form.
In a forthcoming paper \cite{AW} we will show that at incommensurate doping the
phase difference between order parameters in the points (1,2) and (3,4) is the
twice the angle $\theta$ between vectors ${\bf r}_{1,2}$ and ${\bf 
r}_{3,4}$. 

2. Let us start with a general comment about Josephson tunneling in the
presence of orthogonality catastrophe. Let one side of the junction be a BCS
superconductor with a small gap and phase $\phi_0$. Then the standard
Ambegaokar-Baratoff formula gives for the Josephson current
$I\sim {\rm Im}\, e^{-i\phi_0}\sum_{p_n}\int_0^\infty 
F(p_n,\omega)d\omega/\omega$ in terms of the $F$-function of a topological
superconductor at momentum $p_n$ normal to the surface, $F(p,\omega) =  2\pi
i\sum \langle N|c_{p\uparrow}|N+1;p\rangle 
 \langle N+1;p|c_{-p\downarrow}|N+2\rangle  \delta(\omega-\epsilon_{p})$.
In BCS theory the main contribution to the integral comes from intermediate
states with an energy of the order of a gap, i.e. a pair is destroyed while tunneling.  In
a topological superconductor the ground states $|N\rangle $  and low energy states
$|N+1\rangle $ have different topological charges and are orthogonal to each other. Their
overlap vanishes in the macroscopical system. The $F$-function remains nonzero due to a
small  contribution of infinitely many states with energy much larger than the gap. As
a result, $F$ decays slower than $\omega^{-1}$ (in fact
$F(p,\omega)\sim\omega^{-1/2}$), so that $\int dp F(p,\omega)$ 
is not very small at large
$\omega$. Thus, we conclude that the tunneling amplitude is given by the the equal time
two-particle matrix element (\ref{dts}) -  in contrast to BCS,
electron-pairs remain intact while tunneling (for a similar
phenomenon see \cite{AndersonOrt}).

3. A number of successive steps \cite{IS,WTS2,KWRaman} have been made over last years
towards the derivation of the topological mechanism from an electronic model with an
infinite on-site  repulsion
\begin{equation}
\label{tJ}
	H = \sum_{\langle ij\rangle } t_{ij} { c}_{i\sigma}^\dagger
{ c}_{j\sigma} +{ J}_{ij} {\bf S}_{ i}\cdot {\bf S}_{j} \,;\;\;\;
{c}_{i\sigma}^{+} {c}_{i\sigma}^{\phantom +}\ne 2.
\end{equation}
Below we sketch such a derivation and add some new features to take into account spin
correlations. The $t-J$ model does not have distinct scales to isolate the physics of
topological fluids. To capture the physics of interest we employ an {\it adiabatic
approximation}, i.e. we treat a hole's motion in a slowly varying spin background
\cite{P}.

A single hop of a hole destroys the short range antiferromagnet order. However, two
consecutive hops and a spin flip bring the antiferromagnet in order. As a result, the spin
configuration remains approximately unchanged, only in second (even) order in the
hopping process. In order to treat spins adiabatically we must first integrate out these
virtual processes, so that holes  remain on the same sublattice (a Schrieffer-Wolf like
transformation) \cite{KWRaman,WTS2}.

In second order of perturbation theory in $t$ the hopping Hamiltonian is
$$
	{\cal H}= \sum 
\{c_\sigma^{\dagger}(a)t_{\sigma\sigma^{\prime}}
(a, a^{\prime}) c_{\sigma^\prime}(a^{\prime})    \nonumber \\
 + c_\sigma^{\dagger}(b) t_{\sigma\sigma^{\prime}}(b,b^{\prime})
c_{\sigma^\prime} (b^{\prime})\}
$$
where the sum runs over neighbors $(a,a')$ of sublattice A and $(b,b')$ of  B on a
square lattice. The variable hopping amplitudes, say
$t_{\sigma\sigma^{\prime}}(a,a')$ depend on the spin configuration and are superpositions
of chiralities over paths connecting points $a,a'$:
\begin{eqnarray}
\label{ch}
t_{\sigma\sigma^{\prime}}(a,a')&=&{\bar t}\sum_b
W_{\sigma\sigma^{\prime}}(a,b,a'),\\
W_{\sigma\sigma^{\prime}}( a, b, a')
&=&(\frac{1}{2}+\mbox{\boldmath $\sigma$}\cdot{\bf S}_a)
(\frac{1}{2}+\mbox{\boldmath $\sigma$}\cdot {\bf  S}_b)
(\frac{1}{2}+\mbox{\boldmath $\sigma$}\cdot{\bf  S}_{a'})\nonumber
\end{eqnarray}
where the overall scale for hopping is ${\bar t}\sim t^2/J$. Now the hopping
Hamiltonian is ready for the adiabatic approximation.

First we must find a static spin configuration, i.e. the  amplitudes $t(a, a^{\prime})$ to
minimize the energy. Their modulus is determined by the competition between electronic
and magnetic energies and gives the overall scale of the model. As far as the phase of the
hopping amplitudes is concerned we assume that it is determined by the electronic
energy alone \cite{P}. The {\it flux hypothesis} \cite{IS} suggests that, in the leading
order in doping, the energy achieves its minimum if the chiralities along contours
$a\rightarrow a+e_i \rightarrow a+2e_i$ are equal in all directions, while chiralities
along two different paths connecting sites on the diagonal of a crystal cell have a
different sign (the principal of maximal interference) and cancel the amplitude for
diagonal hopping:\\
$\langle W({\bf r},{\bf r}+{\bf e}_x,{\bf r}+{\bf e}_x+{\bf e}_y)+W({\bf
r},{\bf r}+{\bf e}_y,{\bf r}+{\bf e}_x+{\bf e}_y)\rangle =0,$\\
$\langle t({\bf r},{\bf r}+{\bf e}_x+{\bf e}_y)\rangle =0,\;\;\;\;\langle t({\bf
r},{\bf r}+2{\bf e}_i)\rangle ={1/m}$.

The Fermi-surface of the mean field state consists of four pockets around
Dirac points ${\bf  k}_{f}\equiv k_{\pm,\pm} = (\pm\frac{\pi}{2},
\pm\frac{\pi}{2})$, so we decompose electron operator  onto four smooth movers
$c_\sigma({\bf r}) = \sum_{k_f}c_{\sigma, \pm\pm}({\bf r}) e^{i{\bf 
k}_{f}{\bf r}}$. In what follows we refer to the smooth functions $c_{\sigma,
\pm\pm}({\bf r})$ as to the {\em continuum part}  and to the factors $e^{i{\bf 
k}_{f}{\bf r}}$ as to the {\em lattice part} of the fermion operator $c_\sigma({\bf
r})$.

In this basis, the mean field Hamiltonian can be written in a continuum limit as the
square of Dirac operator $H=D^2=(\alpha_x i \partial_x + \alpha_y i \partial_y )^2$
where the $4\times 4$ Dirac matrices $\{\alpha_x,\alpha_y\}=0$ act in the space labeled
by   $(\pm\pm)$. The choice of these matrices (gauge freedom) corresponds to a
relabeling of the Dirac points and is limited by the symmetry group of the Fermi surface
(there are only four different gauges). We choose them to be
$\alpha_x=\tau_3\otimes\tau_3,
\alpha_y=\tau_1\otimes\tau_3$, where the first Pauli matrix $\tau$ acts on the first (x)
$\pm$ label and the second acts on the y $\pm$ label. They correspond to the Landau gauge
on the lattice with hopping amplitudes 
$\alpha_x\equiv\alpha ({\bf r}, {\bf r}+{\bf e}_x)=1, \alpha_y\equiv\alpha ({\bf r},
{\bf r}+{\bf e}_y)=(-1)^x$.

Now we are ready to take into account smooth fluctuations of the phase of the hopping
amplitudes (fluctuations of moduli are not that important). The most effective way to do
this is to introduce a non-gauge-invariant operator $\psi_\sigma$ to describe
charge motion and $U(2)$ gauge field ${\bf  A}+{\bf \cal A}\cdot\mbox{\boldmath $\sigma$}$
to describe  fluctuations of spin chirality $W_{\sigma,\sigma'}\sim 
\exp{i(F+{\bf \cal F} \cdot\mbox{\boldmath $\sigma$}_{\sigma,\sigma'})}$, where $F$
and ${\bf {\cal F}}$
are fluxes of $U(1)$ and $SU(2)$ gauge fields. We find the hopping Hamiltonian to be
Pauli operator:
\begin{eqnarray}
\label{Pauli}
{\cal H} =
\frac{1}{2m}\psi^\dagger\{(i{\bf \nabla}-{\bf  A}-{\bf \cal
A}\cdot\mbox{\boldmath $\sigma$})^2 + \Gamma (  F+{\bf \cal
F}\cdot\mbox{\boldmath $\sigma$})\}\psi
\end{eqnarray}
The  second term with $\Gamma = -i\alpha_x\times\alpha_y$ describes the
diagonal hopping due to the fluctuations of chirality.

5. The perturbative vacuum (where the gauge field ${\bf  A}$  is
small), is unstable when we start to dope the system. The energy achieves
its minimum if the Abelian part $F$ of the flux  (the topological
charge of magnetic solitons) is equal to the number of dopants. The reason
for this is that the non negative hopping Hamiltonian (\ref{Pauli}) in the
presence of a static flux  acquires  states with zero energy \cite{AharonovCasher}. The
number of zero modes is twice the number of flux quanta and the density of particles
occupying zero modes is adiabatically coupled to the flux: $\rho(r)=2\frac{F}{2\pi}$.
The  wave functions (non gauge invariant) of zero modes are $\Phi(r)= e^{i\phi-\Gamma
\chi} g$ where $A_z = \partial_z (-i\phi+\chi)$
and $g$ (the lattice part of the zero mode) obeys $\Gamma g=-g$. If the flux is directed
up there are two solutions
\begin{eqnarray}
	g_{A(B)} & = &  e^{i{\bf  k}_{++}\cdot{\bf  r}}
\pm e^{i{\bf  k}_{--}\cdot{\bf  r}}
+i(e^{i{\bf  k}_{-+}\cdot{\bf  r}}\pm e^{i{\bf  k}_{+-}\cdot{\bf  r}})
	\label{g2bb}
\end{eqnarray}
which are chosen, such that $g_A=0$  on sublattice B and $g_B=0$  on sublattice A.

While doping, electrons want to create and occupy zero mode states to
minimize their energy. This effect competes with the magnetic energy of the
flux. Below we assume that the gain in electronic energy wins this
competition \cite{AW}. As a result two electrons with opposite spins may
occupy the same zero mode state. Once zero mode states are occupied, the
interaction between them lifts the degeneracy, so that the zero mode states
form a narrow band. In a singlet state the band is always completely filled
and is detached from the rest of the spectrum, so that the chemical
potential lies in a gap $\Delta_0$. This results in  superconductivity --- 
a density modulation can propagate together with a flux configuration
without dissipation, while the electronic spectrum has a gap at the Fermi
surface.

A short range antiferromagnetic interaction would suggest that electrons with spin
up (down) spend most of their time on sublattice A(B), so that for low energy
states, we may identify the spin and the sublattice and to project onto
zero mode states. Thus we obtain the $U(1)\times U(1)$ anyon model
\begin{eqnarray}
\label{A}
{\cal H}&=&\frac{1}{2m}\psi^\dagger\{(i{\bf \nabla}-{\bf  A}-{\bf \cal
A}^3\sigma_3)^2\}\psi+2A_0
\frac{F}{2\pi}+\frac{1}{2\lambda}({{\cal F}^3})^2\nonumber\\
&&[A_x(r),A_y(r)]=i\frac{2\pi}{2}\delta (r-r').
\end{eqnarray}
The last term with a phenomenological constant $\lambda$ is added to
implement the magnetic part of the model. It induces an attraction between
particles with opposite spins and determines the gap
$\Delta_0$ and $T_c$.

Here we do not calculate the structure of this narrow band but rather
concentrate on the distribution of electronic spin within the band. It is
governed by the $SU(2)$ part of the flux. In the "unitary" gauge, where
${\cal F}$ is directed along the third axis ${\bf{\cal F}}={\cal F}^3$,  the
spin density is
\begin{equation}
\label{3}
\rho^3(r)\equiv{1}/{2}c^{\dagger}\sigma^3c ={{\cal F}^ 3}/{2\pi}.
\end{equation}
An important supplement to this theory is the identification of the original
electronic operator $c_\sigma$ on the lattice and the non gauge invariant
$\psi_\sigma$ in the continuum. The gauge invariant electron operator creates a
hole plus the flux of the Abelian gauge field attached to it, i.e. it consists of a
product of $\psi_\sigma$ and the vertex operator $V$, which unwinds the
Abelian gauge field:   ${\cal H}({\bf A})V=(V^{*})^{-1}{\cal H}(0)$. It is
$V(z)=\exp\{-\int^r{A}_z(z') dz'\}$, where $z=x+iy$ is  a holomorphic coordinate relative
to the crystal axes. The vertex operator creates flux $[V(r),
F(r')]=\pi V(r)\delta(r-r')$. Thus
$c_\sigma(r)= V(r)\psi_\sigma(r)U(r)g_\sigma(r)$ where the factor
$U(r,C)=\prod_C\alpha_i$ is the lattice part of the Dirac tail along some contour. It
 makes the wave function of the zero mode gauge invariant. The Hamiltonian
(\ref{A}) together with the correspondence between continuum  and  lattice fields is
the field theory for a doped Mott insulator on a bipartite lattice.

6. We now proceed with matrix elements. Let us add two particles in a
singlet state into a state with $N$ particles. The ground state with $N+2$
particles consists of two extra electrons in the zero mode state and also a
proper redistribution of the flux ${\cal F}^3$:
\begin{equation}\label{N}
|N+2\rangle =\varepsilon_{\sigma\sigma'}\int drdr'
\mu(r,r')c_{\sigma}^\dagger (r) c_{\sigma}^\dagger(r')|N\rangle
\Psi(r,r')  
\end{equation}
Operator $\mu(r,r')$  creates the flux of the
${\cal A}^3$ field according to (\ref{3}):
$[{\cal F}^3(u),\mu(r,r')]=\pi\mu(r,r')(\delta(r-u)-\delta(r'-u))$, and
$\Psi(r,r')$ is the wave function of a singlet in the zero mode state.
A solution for the flux operator is $\mu(r,r')=V_3(r)V_3^{-1}(r')$ where
$V_3(z)$ is the vertex operator
\begin{equation}
\label{vv}
V_3(z)\sim e^{-\int^z{A^3}_z(z') dz'}=
e^{\int \ln\frac{(z-z')}{L}\rho_3(z')dz'd\bar z'}
\end{equation}
and $L$ is the size of the system. The meaning of this result is simple. The
form of operator $\mu$  suggests that in the  spin
singlet state electrons with spin $\sigma$ see $+2\pi$ flux attached to the
other electrons with the same spin while  $-2\pi$ attached to particles with
the opposite spin \cite{Girvin}. This can be illustrated by the operator
algebra:
\begin{eqnarray}
\label{OPA}
V_3(z)c_{\uparrow,\downarrow}(z')   &\sim&
(\frac{z-z'}{L})^{\pm 1/2}c_{\uparrow,\downarrow}(z')V_{3}(z)
\end{eqnarray}
By means of  (\ref{OPA}) we obtain 
\begin{eqnarray}
\label{two particles}
\Delta(z,z')=\frac{a}{(z-z')}\Psi(z,z')\langle N|\mu(z,z'))|N\rangle .
\end{eqnarray}
The last factor here is merely a constant while two others contribute to the
angle dependence.

Let us first note that the size of the system  dropped out and is replaced by a
short distance cutoff $a$. This shows that the overlap of the ground states with
$N$ and $N+2$ particles does not vanish in a macroscopic system \cite{Talstra}. In
contrast, an attempt to insert a single electron into the system leaves a zero mode
unfilled and  creates a non-singlet excitation. This leads to an orthogonality
catastrophe: Due to the operator algebra (\ref{OPA}) the matrix element
$
\langle N|c_\sigma|N+1\rangle \sim \langle N|c_\sigma(0)\int dz
V_\sigma(z)c^\dagger_\sigma(z)|N\rangle \sim\frac{a}{L}
$
vanishes as $L\rightarrow 0$.

\begin{figure}
\centerline{\psfig{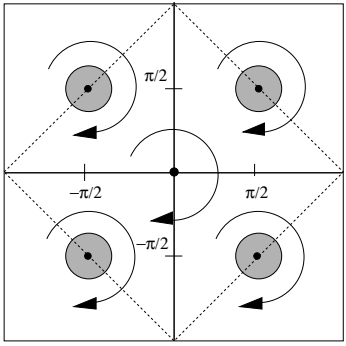}}
\vspace{0.2cm}
\narrowtext
\caption{The order parameter in momentum space: 
$\Delta({\bf k})$ consists of four similarly oriented
unit vortices around  each Fermi point $(\pm\frac{\pi}{2},\pm\frac{\pi}{2})$. The
phase of each vortex is relative to the direction of the ${\bf k}_f$ as in eq.
(\protect\ref{popo}).  It changes the sign under $90^o$ rotation
$\Delta(-k_y,k_x)=-\Delta(k_x,k_y)$.}
\label{Fig1}
\end{figure}

7.  The  two particle wave function $\Psi({\bf r},{\bf r}')$ in (\ref{N}) consists of
a lattice part and the smooth BCS wave function of the pair
$\Psi({\bf r},{\bf r}')=\Psi_{\mbox{\it{lattice}}}({\bf r},{\bf r}')\Delta_{BCS}({\bf
r}-{\bf r}')$
$$\Delta_{BCS}({\bf r})\sim \int dk e^{i{\bf  k}\cdot{\bf  r}}
\frac{\Delta_0}{\sqrt{(\epsilon_{\bf k}-\mu)^2+\Delta_0^2}}$$ where  ${\bf k}$ 
is 
relative to minima ${\bf k}_{\pm\pm}$ of the mean field spectrum $\epsilon_k$, and
$\Delta_0$ is a gap which separates the narrow band from the spectrum. The lattice
part of the wavefunction is $\Psi_{\mbox{\it lattice}}({\bf r},{\bf r}')=U({\bf
r},C_r)U({\bf r}',C_r')g_A({\bf r})g_B({\bf r}')$ depends on two strings (contours)
ended in points ${\bf r}$ and ${\bf r}'$. Fluctuations of the string are physical
excitations of the pair  (not an artifact of the approach). In the commensurate case,
strings fall in four groups within which $\Psi_{\mbox{\it lattice}}$ is the same.
These groups correspond to the states with pairing from different Fermi points ${\bf
k}_f$ and ${\bf k}'_f$, i.e. to a pair with a total momentum
${\bf P}={\bf k}_f+{\bf k}'_f=(0,0)\equiv (\pm\pi,\pm\pi), (\pm\pi,0),(0,\pm\pi)$.  A
particular string of the wave function  with momentum ${\bf P}=0$ can be chosen as two
contours following each other from some reference point up to the point ${\bf r}=(x,y)$
and then a single string along the $y$-axis to $(x,y')$ and finally to the point
${\bf r}'=(x',y')$ along the
$x$-axis. In the chosen gauge  this factor is $U({\bf
r},C_r)U({\bf r}',C_r')=(-1)^{x(y-y')}$. Then the order parameter (\ref{dts}) is
translational invariant.  Combining this and (\ref{g2bb}) we obtain:
\begin{equation}
\label{kl}
\Delta({\bf R})
\sim
\frac{\sin\frac{\pi}{2}(X+Y)+i\sin\frac{\pi}{2}(X-Y)}
{X+iY}\Delta_{BCS}({\bf R})
\end{equation}
where ${\bf R}={\bf r}-{\bf r}'\equiv (X,Y)$.
The numerator of this expression is a discrete analog of the continuous
holomorphic function in the denominator. Under $\pi/2$ rotation it produces
the factor $e^{-i\pi/2}$. Another $e^{-i\pi/2}$ factor is produced by the
continuum part. Both phases add to $e^{-i\pi}=-1$, so that the tunneling
amplitude belongs to an irreducible d-representation  of the crystal
group $\Delta(-Y,X)=-\Delta(X,Y)$. It is instructive to look at the tunneling
amplitude in momentum space. It is:
\begin{eqnarray}
\label{popo}
\Delta({\bf k}) = \sum_{{k}_f}
e^{-i\arg({\bf k}_f)-i\arg({\bf k}-{\bf k}_f)} f(|{\bf k}-{\bf k}_f|),
\end{eqnarray}
where $\arg({\bf k}_f)=\frac{\pi}{4}, \frac{3\pi}{4}, \frac{5\pi}{4},
\frac{7\pi}{4}$ and $f(p)$ is a smooth function. The tunneling
amplitude  consists of two vortices -- one in the center
of the Brillouin zone (lattice part) while another is at a Fermi point
(Fig.\ref{Fig1}).
\vspace{-0.2cm}

8. Eq.(\ref{popo}) suggests an interesting generalization to
the incommensurate case (optimal doping) where the Fermi surface is a
  simply connected curve rather than four Fermi pockets. We
 rewrite (\ref{kl}) approximately as
$$
\Delta({\bf r})=\frac{e^{-i\arg({\bf r})}}{|{\bf r}|}\int dk e^{i{\bf  k}\cdot{\bf 
r}}e^{-i\arg({\bf k})}
\Delta_{BCS}({\bf k})
$$
Here ${\bf k}$ is relative to the center of the Brillouin zone and 
the factor $\arg({\bf k}_{\pm\pm})$ is replaced by
$\arg({\bf k})$. It is possible, because $\Delta_{BCS}({\bf k})$ set the integral
 to the Fermi surface. At small doping it is shown in Fig.\ref{Fig1}.
Away from half-filling
 $\epsilon_{\bf k}-\mu\sim v_f(|{\bf k}|-k_f)$. Thus we  obtain a
"tomographic" representation of the order parameter \cite{AL}:
$$\Delta({\bf k})= e^{-i2\arg({\bf k})}\int d{\bf  k}_f
\Delta_{BCS}({\bf k}-{\bf k}_f-{\bf  q})D_{{\bf  k}-{\bf  k}_f}(q)d{\bf  q}$$ where
the propagator $D_{\bf p}({\bf q})=|{\bf  q}|^{-1}\exp{(i\hat{{\bf p}{\bf  q}})}$ is a
holomorphic function of ${\bf q}$. Here 
$\hat{{\bf p}{\bf  q}}$ is an angle between ${\bf p}$ and ${\bf  q}$ and the integral
$d{\bf  k}_f$ goes over the Fermi surface.

This is the main result of this paper. It clarifies the physics of
a topological superconductor. In contrast to BCS, an  electron with momentum
${\bf  k}$ close to ${\bf  k}_f$ emits soft modes of density modulation
with the propagator $D_{{\bf  k}-{\bf   k}_f}(q)$. As a  result: (i) the
ground states differing by an odd number of particles are orthogonal; (ii) the BCS
wave function is dressed by soft density modes. This is analogous to 1D physics and
{\it bremsstrahlung} of QED. The new features are: (i) emission of the soft mode is
forward; (ii) the phase of the
matrix element of the soft mode is the angle relative to the Fermi momentum. 

9. Although the order parameter (\ref{popo}) forms a d-representation, its form and
the physics behind it are drastically different from the conventional d-wave  $\cos
k_x -\cos k_y$. Nevertheless, it seems premature to speculate on an observable
difference, until interlayer tunneling is taken into account. This is crucial,
since the sign of parity breaking alternates between odd and even layers
\cite{LZL,WTS2}, and a realistic junction averages over many layers. In a
hypothetical monolayer tri-crystal experiment one would expect the 
trapped flux to be an integer (in contrast to half-integer for conventional
d-wave \cite{Tsuei}).

We would like to thank A. Larkin, K. Levin, B. Spivak and J. Talstra  for
numerous discussions and L. Radzihovsky for collaboration on the initial
stage of this project. This work was supported by MRSEC NSF Grant DMR 9400379 and NSF
Grant DMR 9509533. AGA also thanks Hulda B.\ Rotschild Fellowship for support.

\end{document}